# $^8$Li(n,γ)$^9$Li REACTION AT SUPERLOW ENERGIES

S. B. Dubovichenko[1,†], A. V. Dzhazairov-Kakhramanov[1,‡]

[1]*V. G. Fessenkov Astrophysical Institute "NCSRT" NSA RK, 050020, Observatory 23, Kamenskoe plato, Almaty, Kazakhstan*
[†]*dubovichenko@mail.ru*, [‡]*albert-j@yandex.ru*

**ABSTRACT**

Within the framework of the modified potential cluster model with the state classification of nuclear particles according to Young tableaux the possibility of describing the available experimental data for the total cross sections of the neutron radiative capture on $^8$Li at thermal and astrophysical energies was considered.

*Subject Keywords*: Nuclear astrophysics; primordial nucleosynthesis; light atomic nuclei; low and astrophysical energies; radiative capture; thermonuclear processes.

## 1. Introduction

Earlier, in our papers (Dubovichenko & Dzhazairov-Kakhramanov 1994, 1997), the possibility of describing Coulomb form-factors of lithium based on two-body potential cluster model (PCM) (Dubovichenko 2012a; Dubovichenko et al. 1990; Neudatchin, Sakharuk & Dubovichenko 1995) was shown. Such model takes into account forbidden states (FSs) (Dubovichenko et al. 1990; Neudatchin, Sakharuk & Dubovichenko 1995; Dubovichenko 1995, 2012b, 2013a; Dubovichenko & Dzhazairov-Kakhramanov 1990) in intercluster potentials, which were used by us in the papers (Dubovichenko & Zhusupov 1984a, 1984b). Furthermore, in the potential cluster model with tensor forces (Dubovichenko 1998; Kukulin, Pomerantsev, Cooper & Dubovichenko 1998), the possibility of the correct reproduction of almost all characteristics of $^6$Li including its quadrupole moment was demonstrated. Finally, in the papers (Dubovichenko 2010, 2011a, 2011b, 2012a, 2012c, 2012d, 2013b, 2013c, 2013d; Dubovichenko et al. 1990, 2014; Neudatchin, Sakharuk & Dubovichenko 1995; Dubovichenko & Dzhazairov-Kakhramanov 2009a, 2009b, 2012a, 2012b, 2012c, 2013, 2014; Dubovichenko & Burkova 2014; Dubovichenko & Uzikov 2011; Dubovichenko Dzhazairov-Kakhramanov & Afanasyeva 2013; Dubovichenko, Dzhazairov-Kakhramanov & Burkova 2013) the possibility of describing the astrophysical *S*-factors or the total cross section of the radiative capture reactions in the p$^2$H, n$^2$H, p$^3$H, p$^6$Li, n$^6$Li, p$^7$Li, n$^7$Li, p$^9$Be, n$^9$Be, p$^{10}$B, n$^{10}$B, p$^{11}$B, n$^{11}$B, p$^{12}$C, n$^{12}$C, p$^{13}$C, n$^{13}$C, p$^{14}$C, n$^{14}$C, n$^{14}$N, p$^{15}$N, n$^{15}$N, n$^{16}$O, $^2$H$^4$He, $^3$He$^4$He, $^3$H$^4$He, and $^4$He$^{12}$C channels was shown at thermal and astrophysical energies. Calculations of these processes were carried out on the basis of the modified variant of the PCM (MPCM) with FSs and the state classification according to Young tableaux, which was described in papers (Dubovichenko 2013d, 2014; Dubovichenko & Dzhazairov-Kakhramanov 2012b, 2012c, 2013, 2014; Dubovichenko & Uzikov 2011; Dubovichenko, Dzhazairov-Kakhramanov & Afanasyeva 2013; Dubovichenko, Dzhazairov-Kakhramanov & Burkova 2013).

Well-defined success of the MPCM at the description of the total cross sections of this type can be explained by the fact that the potentials of the intercluster interaction in the continuous spectrum are constructed on the basis of the known elastic scattering phase shifts or structure of the spectrum levels of the final nucleus, and for the discrete spectrum - on the basis of the main characteristics of the bound states (BSs) of such nuclei. These potentials are constructed taking into account the classification of the cluster states according to Young tableaux (see, e.g., Neudatchin Smirnov 1969), which give the possibility to determine the presence and quantity of the FSs in each partial wave. This means that the number of wave function (WF) nodes in such cluster systems (Dubovichenko 2014) are also determined.

The potential parameters of the BSs at the given number of allowed (ASs) and forbidden states in the given partial wave states are fixed quite unambiguously according to the binding energy, the nucleus radius and the asymptotic constant (AC) in the considered two-body channel. The ambiguities of these parameters are caused by the extract error, for instance, the AC from the experimental data. Potential scattering parameters are also fixed quite clearly according to the scattering phase shifts or resonances in the spectra of the final nucleus, for instance, $^9$Li in the n$^8$Li channel, if the number of FSs and ASs is given in each partial wave (Dubovichenko 2014). Ambiguities of such potentials are caused by the scattering phase shift error or the error in determination of the energy and the width of the resonance level.

In fact, these additional conditions on the number of FSs, which are imposed on the parameters of the partial potentials, allow to get rid of a discrete ambiguity that is well known in the optical model (Hodgson 1963). Moreover, requirements for description of the BS AC or resonance in the nuclear spectrum at the positive energies relativity to the cluster channel threshold allow one to get rid from the continuous ambiguity of such potentials (Dubovichenko 2014). In other words, if the number of FSs and ASs are known in the given partial wave, then the potential parameters for the binding energy, the AC or parameters of the resonance are fixed almost unambiguously and depend just on errors of these values.

Furthermore, such potentials allow carrying out the calculations of some basic characteristics of interaction of the considered particles in the elastic scattering processes and reactions. For instance, it can be the astrophysical *S*-factors of the radiative capture reactions (Angulo et al. 1999) or the total cross sections of these reactions (Adelberger 2011). Including radiative capture cross sections for some thermonuclear reactions at astrophysical and thermal energies, which are considered in our previous papers.

Therefore, continuing study of the radiative capture processes (Dubovichenko 2012a, 2014), let's consider the n+$^8$Li→$^9$Li+γ reaction at thermal and astrophysical energies. Some other calculation results of the total cross section of this reaction, based on different models and methods, which describe existing experimental data (Zecher 1998), can be found, for instance, in papers (Zecher 1998; Banerjee Chatterjee & Shyam 2008). Let us note right now that we could find in (Nuclear Reaction Database (EXFOR), 2013) just experimental data of work (Zecher 1998) and not so many theoretical calculations of the cross sections, for example, (Zecher 1998; Banerjee Chatterjee & Shyam 2008; Bertulani 1999) and some other, results of which differ significantly from each other.

## 2. Astrophysical Aspects

The n+$^8$Li→$^9$Li+γ reaction can present a significant astrophysical interest, because it includes into one of the variants of chain of primordial nucleosynthesis processes of the Universe and thermonuclear reactions in type II supernovae (Banerjee Chatterjee & Shyam 2008). In the process of such fusion after the nuclear formation with $A$ = 7 the further way of formation of the elements with $A$ = 11 and higher can occur, for example, with help of the $^7$Li(n,γ)$^8$Li(α,n)$^{11}$B reaction or due to less likely branch $^7$Li(α,γ)$^{11}$B (Boyd et al. 1992; Gu et al. 1995). However, how it was shown in paper (Malaney & Fowler 1988), the neutron capture on $^8$Li, i.e., the n+$^8$Li→$^9$Li+γ reaction can reduce the amount of these nuclei in the given chain of fusion and this reduction can reach the value of 50%. In other words, there are at least two variants of the primordial synthesis of elements with mass of $A$ = 11-12 and further. Notably, such synthesis can take place not only due to the chain $^7$Li(n,γ)$^8$Li(α,n)$^{11}$B(n,γ)$^{12}$B(β$^+$)$^{12}$C, but with another way, for example, according to the reactions $^7$Li(n,γ)$^8$Li(n,γ)$^9$Li(α,n)$^{12}$B(β$^+$)$^{12}$C (Guimarães et al. 2006). The considered here reaction $^8$Li(n,γ)$^9$Li is the basis of this chain. It is possible to "produce" heavy isotopes using the $r$-process for supernovae stars of type II, immediately after the collapse stage. Particularly, at the beginning of the expansion stage the gap mass of elements at $A$ = 8 can be eliminated by the reactions of the type α+α+α→$^{12}$C or α+α+$n$→$^9$Be (Woosley et al. 1994). However, the overcoming of such a mass gap is possible through the chain of reactions $^4$He(2n,γ)$^6$He(2n,γ)$^8$He(β$^-$)$^8$Li(n,γ)$^9$Li(β$^-$)$^9$Be (Görres et al. 1995; Efros et al. 1996) on this stage, where the considered here process is also presented.

The problem in the investigation of the reaction $^8$Li(n,γ)$^9$Li is that the direct experimental measurement of the cross sections is impossible, because the half-life of $^8$Li is too small – 838 ms (Guimarães et al. 2006). The microscopic calculations allowed us to get the rate of this reaction (Bertulani 1999; Rauscher et al. 1994; Descouvemont 1993), however such results differ about an order of magnitude. For instance, in the comparative table of the results for the neutron capture reaction rate on $^8$Li from paper (Horváth et al. 2003) the values from 5300 to 790 cm$^3$s$^{-1}$mole$^{-1}$ are given. Therefore, lately, the experimental efforts were carried out for study the reaction $^8$Li(n,γ)$^9$Li with the help of indirect approaches, for example, Coulomb dissociation (Kobayashi et al. 2003) and (d,p) transfer reactions (Guimarães et al. 2006; Guo et al. 2005). Consequently, the total cross sections of the neutron capture on $^8$Li were obtained (Zecher 1998), and then we use them further for the comparison with our theoretical calculations. Great ambiguities in the study of this reaction by different methods make it interesting as a subject for study. Therefore we will consider the neutron capture on $^8$Li reaction in the frame of the MPCM (Dubovichenko 2014) and will define how the criteria of this model can properly describe the total cross sections for the neutron capture on $^8$Li at thermal and astrophysical energies, namely from 25.3 meV (1 meV = 10$^{-3}$ eV) to 1.0 MeV.

## 3. Classification of n$^8$Li states according to Young tableaux

For $^8$Li, as well as for $^8$Be, accept the Young tableau {44}, because we have {44} + {1} = {54} + {441} for the n$^8$Li system. It is clear, that in the state with $L$ = 0, which we will need in the future, the FS with the tableau {54} is present. Allowed state corresponds to the configuration {441} at $L = 1$ – orbital angular momentum is determined by the Eliot's rule (see, e.g., Neudatchin Smirnov 1969). The first of the obtained tableaux is forbidden, as far as there can't be five nucleons in the s-shell, the second is allowed and compatible with the orbital angular momentum $L = 1$, to which corresponds the ground state (GS) of $^9$Li in the n$^8$Li channel (Tilley et al. 2004).

Thus, limited by the lower partial waves with the orbital angular momentum $L = 0,1$ it can be said that for the n$^8$Li system (for $^8$Li is known $J^\pi, T = 2^+, 1$ (Tilley et al. 2004)) in the potentials of the $P$ waves only the AS are present, while there are FS in the $S$ waves. The state in $P_{3/2}$- wave corresponds to the GS of $^9$Li with $J^\pi, T = 3/2^-, 3/2$ and located at the binding energy of the n$^8$Li system of -4.0639 MeV (Tilley et al. 2004). Some n$^8$Li scattering states and BSs can be mixed by spin with $S = 3/2$ ($2S + 1 = 4$) and $S = 5/2$ ($2S + 1 = 6$). However, here we will consider that the GS of $^9$Li in the n$^8$Li channel most probably is the $^4P_{3/2}$ level (in notations $^{(2S+1)}L_J$), although in both spin states for $L = 1$ the total angular moment $J = 3/2$ is possible.

In this case, we do not have the full tables of products of Young tableaux for a system with the number of particles greater than eight (Itzykson & Nauenberg 1966), which were used by us earlier for the such calculations (Dubovichenko 2012a, 2014). Therefore, the obtained results should be considered only as a qualitative estimation of possible orbital symmetries in the ground state of $^9$Li in the n$^8$Li channel. However, exactly on the basis of such classification it was able to quite reasonably explain the available experimental data on radiative capture of neutrons and other particles for a wide range of nuclei of 1$p$-shell with masses from 6 to 16 (Dubovichenko 2012a, 2014). Therefore, here we will use the classification of cluster states by orbital symmetry, which leads us to a certain number of FSs and ASs in the partial intercluster potentials; it means that to a certain number of nodes of the relative motion wave function of clusters – in the given case, neutron and $^8$Li. Even as a qualitative estimation of orbital symmetries allow to determine the presence of the FS in the $S$ wave and absence of the FS for the $P$ states. Exactly the same structure of FSs and ASs in the different partial waves allows one to build the potentials of intercluster interaction that are necessary for calculations of the total cross section of the considered radiative capture reaction.

## 4. Structure of states in the n$^8$Li channel

Now let's consider the structure of the excited and resonance states of $^9$Li in the n$^8$Li channel. Furthermore we will take into account the first excited state (FES), which is bound in the n$^8$Li channel, and the first resonant state (FRS) of $^9$Li, because only for them the momenta and parities (Tilley et al. 2004) are known. The FES of $^9$Li with $J^\pi = 1/2^-$ is located at the energy of 2.691(5) MeV relative to the GS or -1.3729(5) MeV relative to the threshold of the n$^8$Li channel. This state is a quartet $^4P_{1/2}$ level and its potential will have one bound AS.

The first resonance state is located at 4.296(15) MeV relative to the GS or 0.2321(15) MeV relative to the threshold of the n$^8$Li channel. For this level the momentum $J^\pi = 5/2^-$ is given (Tilley et al. 2004), this allows to take $L = 1$ for it, i.e., to consider it as a quartet $^4P_{5/2}$ resonance at 0.261(17) MeV in the laboratory system (l.s.). Its width equals $\Gamma_{cm} = 100(30)$ keV is given in paper (Tilley et al. 2004). It is possible to construct the quite unambiguous $^4P_{5/2}$ potential of the elastic scattering (Dubovichenko 2014) from these data. The ambiguities of its parameters will be caused only by the error of the width of this resonance, because it doesn't contain bound FSs or ASs.

On the basis of these data one can consider that $E1$ capture is possible from the $^4S_{3/2}$ scattering wave to the $^4P_{3/2}$ GS of $^9$Li. As far as the GS with $^4P_{3/2}$ and the FES with $^4P_{1/2}$ we consider as the quartet states then the main contribution will give the transition of the form $^4S_{3/2} \to\, ^4P_{3/2}$. For the radiative capture on the FES the similar $E1$ transition $^4S_{3/2} \to\, ^4P_{1/2}$ is possible. Since the results of the phase analysis of the elastic n$^8$Li scattering could not be found, furthermore several variants of the $^4S$ potential with the bound FS and different widths will be discussed. Such potentials should lead to the $^4S$ phase shifts that close to zero, as in the spectra of $^9$Li there are no $^4S$ resonances. Here, it should be noted that the FS will not be bound, necessarily. The above classification allows one to determine the presence of the FSs, but does not allow one to set whether it is bound.

Furthermore the GS potentials will be constructed for a correct description of the channel binding energy, the charge radius of $^9$Li and its asymptotic constant in the n$^8$Li channel. Therefore, the known values of the asymptotic normalization coefficient (ANC) and the spectroscopic factor $S$, with help of which the AC is found, have a quite big error, the GS potentials will also have several variants with different parameters of width.

The radius equals of 2.327 ± 0.0298 fm was used in further calculations that is given in the database (Chart nucl. shape & size param. (0 ≤ Z ≤ 14) 2015). There is known the value of the radius of 2.2462 ± 0.0315 fm for $^9$Li that is given in the database (Chart nucl. shape & size param. (0 ≤ Z ≤ 14) 2015). Moreover, for example, in paper (Sanchez et al. 2006) for radiuses of these nuclei the next values are found: 2.299(32) fm for the $^8$Li and 2.217(35) fm for the $^9$Li. In paper (Nortershauser et al. 2005) for these radiuses were obtained values 2.30(4) fm and 2.24(4) fm accordingly. All of these data within the error limits quite conform to each other. For the masses of nucleus and neutron the exact values were used: $^8$Li – 8.022487 amu (Nuclear Wallet Cards database ($^8$Li isotope), 2014) and $n$ – 1.00866491597 amu (Fundamental Physical Constants, 2010). The charge radius of the neutron is assumed to be zero and its mass radius equals 0.8775(51) fm coincides with the known radius of proton (Fundamental Physical Constants, 2010).

Data on the spectroscopic factor $S$ of the GS and the asymptotic normalization coefficients $A_{NC}$ (ANC) are given, for example, in paper (Guimarães et al. 2006). Here we also use the relationship

$$A_{NC}^2 = S \times C^2, \qquad (1)$$

where $C$ is the asymptotic constant in fm$^{-1/2}$, which is related to the dimensionless AC $C_W$ (Plattner & Viollier 1981), used by us in the following way: $C = \sqrt{2k_0} C_W$. The dimensionless constant $C_W$ is determined by the expression (Plattner & Viollier 1981)

$$\chi_L(r) = \sqrt{2k_0} C_w W_{-\eta L+1/2}(2k_0 r), \qquad (2)$$

where $\chi_L(r)$ is the numerical wave function of the bound state, obtained from the solution of the radial Schrödinger equation and normalized to unity, the value $W_{-\eta L+1/2}$ is the Whittaker function of the bound state, determining the asymptotic behavior of the wave function and that is the solution of the same equation without the nuclear potential; $k_0$ is the wave number, caused by the channel binding energy $E$: $k_0 = \sqrt{2\mu \frac{m_0}{\hbar^2} E}$; $\eta$ is the Coulomb parameter $\eta = \frac{\mu Z_1 Z_2 e^2}{\hbar^2 k}$, determined numerically $\eta = 3.44476 \cdot 10^{-2} \frac{\mu Z_1 Z_2}{k}$ and $L$ is the orbital angular momentum of the bound state. Here $\mu$ is the reduced mass, and the constant $\hbar^2/m_0$ assumed to be equal 41.4686 fm$^2$, where $m_0$ is the atomic mass unit (amu).

### 5. Methods for calculating of the total cross sections

The total cross sections of the radiative capture $\sigma(NJ,J_f)$ for the $EJ$ transitions in the potential cluster model are given, for example, in papers (Dubovichenko 2012a, 2014; Dubovichenko & Dzhazairov-Kakhramanov 2012b, 2014; Dubovichenko, Dzhazairov-Kakhramanov & Afanasyeva 2013; Dubovichenko, Dzhazairov-Kakhramanov & Burkova 2013) or (Angulo et al. 1999) and have the form

$$\sigma_c(NJ,J_f) = \frac{8\pi K e^2}{\hbar^2 q^3} \frac{\mu}{(2S_1+1)(2S_2+1)} \frac{J+1}{J[(2J+1)!!]^2} A_J^2(NJ,K) \cdot$$

$$\cdot \sum_{L_i,J_i} P_J^2(NJ,J_f,J_i) I_J^2(J_f,J_i),$$

where $\sigma$ is the total cross section of the radiative capture process, $\mu$ is the reduced mass of the particle of the input channel, $q$ is the wave number of the particles of the input channels, $S_1$, $S_2$ are the particle spins in the input channel, $K$, $J$ are the wave number and $\gamma$-quantum moment in the final channel, $N$ are the $E$ or $M$ transitions of $J$-th multipolarity from the initial $J_i$ to the final $J_f$ state of the nucleus.

For the electric orbital $EJ(L)$ transitions ($S_i = S_f = S$) the value $P_J$ has the form (Dubovichenko 2012a, 2014)

$$P_J^2(EJ,J_f,J_i) = \delta_{S_i S_f} [(2J+1)(2L_i+1)(2J_i+1)(2J_f+1)](L_i 0 J 0 | L_f 0)^2 \begin{Bmatrix} L_i & S & J_i \\ J_f & J & L_f \end{Bmatrix}^2$$

$$A_J(EJ,K) = K^J \mu^J \left( \frac{Z_1}{m_1^J} + (-1)^J \frac{Z_2}{m_2^J} \right), \qquad I_J(J_f,J_i) = \langle \chi_f | R^J | \chi_i \rangle.$$

Here $S_i$, $S_f$, $L_f$, $L_i$, $J_f$, $J_i$ are total spins and momenta of particles of the input ($i$) and the final ($f$) channels, $m_1$, $m_2$, $Z_1$, $Z_2$ are masses and charges of particles of the input channel, $I_J$ is the wave function integral of the initial $\chi_i$ and the final $\chi_f$ state, as the relative motion functions, in this case, of neutron and $^8$Li with the interparticle distance $R$.

For the spin part of the magnetic process $M1(S)$ in the used model the following expression is obtained ($S_i = S_f = S$, $L_i = L_f = L$) (Dubovichenko 2014):

$$P_1^2(M1, J_f, J_i) = \delta_{S_i S_f} \delta_{L_i L_f} [S(S+1)(2S+1)(2J_i+1)(2J_f+1)] \begin{Bmatrix} S & L & J_i \\ J_f & 1 & S \end{Bmatrix}^2,$$

$$A_1(M1, K) = i\frac{\hbar K}{m_0 c}\sqrt{3}\left[\mu_1 \frac{m_2}{m} - \mu_2 \frac{m_1}{m}\right], \quad I_J(J_f, J_i) = \langle \chi_f | R^{J-1} | \chi_i \rangle, \quad J=1.$$

Here $m$, $m_1$, $m_2$ are the masses of the nucleus and clusters; $\mu_1$, $\mu_2$ are the magnetic moments of clusters, the remaining denotations are as in the previous expression. The magnetic moment of neutron is $\mu = -1.91304272\mu_0$ (Fundamental Physical Constants, 2010), and for $^8$Li is accurately measured in (Neugart et al. 2008) and equals $1.653560\mu_0$, where $\mu_0$ is the nuclear magneton.

## 6. The n$^8$Li interaction potentials

As usual in (Dubovichenko 2012a, 2014), we use the Gaussian potential with the point-like Coulomb term as the n$^8$Li interaction in each partial wave with the given orbital momentum $L$

$$V(r, L) = -V_L \exp(-\gamma_L r^2). \tag{3}$$

The ground state of $^9$Li in the n$^8$Li channel is the $P_{3/2}$ level and this potential should describe the AC for this channel correctly. In order to extract the constant $C$ in the form (1) or $C_W$ (2) from the available experimental data, let's consider the information about the spectroscopic factors $S$ and the asymptotic normalization coefficients $A_{NC}$. For example, the data that are given in paper (Kanungo et al. 2008) instead of own results, and data for the previous works, which there were relatively small for the considered system. If to highlight from these results the similar ones, i.e., with the close spectroscopic factor values, we can present them in the form of Table 1.

Table 1. The data on spectroscopic factors $S$ for the GS of $^9$Li in the n$^8$Li from works (Kanungo et al. 2008; Wuosma et al. 2005; Li et al. 2005; Guimarães et al. 2007).

| Reaction from what the spectroscopic factor $S$ is obtained | The spectroscopic factor values for the channel n+$^8$Li$_{g.s.}$ | Ref. |
|---|---|---|
| $^2$H($^9$Li,$^3$H) | 0.65(15) | (Kanungo et al. 2008) |
| $^2$H($^9$Li,$^3$H) | 0.59(15) | (Kanungo et al. 2008) |
| $^2$H($^8$Li,$^1$H) | 0.90(13) | (Wuosma et al. 2005) |
| $^2$H($^8$Li,$^1$H) | 0.68(14) | (Li et al. 2005) |
| $^9$Be($^8$Li, $^9$Li) | 0.62(7) | (Guimarães et al. 2007) |
| *Average value* | *0.69* | --- |
| *Limit* | *0.44-1.03* | --- |

Furthermore, in two well-known papers the square of the ANC of the GS is determined – it equals (Dubovichenko & Dzhazairov-Kakhramanov 2012b), where $A^2_{NC} = 0.92(14)$ fm$^{-1}$ [$A_{NC} = 0.96(8)$ fm$^{-1/2}$] and in (Guo et al. 2005) is obtained that $A^2_{NC} = 1.33(33)$ fm$^{-1}$ [$A_{NC} = 1.15(14)$ fm$^{-1/2}$]. Average between them is equal to $A_{NC} = 1.06$ fm$^{-1/2}$ – this value agrees well with the results of the review of paper (Timofeyuk 2013), where for the GS the value of 1.08 fm$^{-1/2}$ with the value of spectroscopic factor from 0.58 to 1.02 (average 0.8) and ab initio spectroscopic factor equals 0.99 (Wiringa 2014) are given. On the basis of the expressions (1) and the average $S$ for the AC of the GS from the Table 1 we find $C = 1.28$ fm$^{-1/2}$, and since $\sqrt{2k_0} = 0.915$, the dimensionless AC, defined as $C_W = C/\sqrt{2k_0}$, is equal to $C_W = 1.39(15)$.

However, if we use the upper value of the $A_{NC}$ that equal to 1.29 from paper (Guo et al. 2005) and the lower – 0.88 from (Guimarães et al. 2006), then at the average $S = 0.69$ for the dimensionless constant $C_W$ we will obtain a wider range of values of 1.43(27). It can be estimated one more interval of AC values by using the given above range for the $A_{NC}$ and the interval for the spectroscopic factor value $S$, which is shown in Table 1. Then for the dimensionless $C_W$ we will get the wider range of values of 1.54(58).

Now we will note that for the ANC of the GS of $^9$Li in (Huang, Bertulani & Guimaraes 2010) at the spectroscopic factor of $S = 0.8$ the value $A_{NC} = 1.12$ fm$^{-1/2}$ was obtained, i.e., $C = 1.25$ fm$^{-1/2}$, which differs slightly from the given above average value of the ANC. Meanwhile, for the FES of $^9$Li in paper (Huang, Bertulani & Guimaraes 2010) the value of 0.4 fm$^{-1/2}$ at $S = 0.55$ was obtained, which leads us to the $C = 0.54$ fm$^{-1/2}$ or to the $C_W = 0.77$ at $\sqrt{2k_0} = 0.698$ in dimensionless form. In the calculations (Nollett & Wiringa 2011) for the GS the value of $A_{NC} = 1.140(13)$ fm$^{-1/2}$ was obtained, and for the FES the value of ANC is equal to $A_{NC} = 0.308(7)$ fm$^{-1/2}$, that in the large agree with all previous values.

The GS potential that allows one to get a dimensionless constant $C_W$ is equal to 1.39 and has the parameters

$$V_{3/2} = 65.788593 \text{ MeV and } \gamma_{3/2} = 0.18 \text{ fm}^{-2}. \tag{4}$$

It leads to the binding energy of -4.063900 MeV with an accuracy of the finite-difference method (FDM) for searching energy used here equals 10$^{-6}$ MeV (Dubovichenko 2012b), $C_W = 1.40(1)$ on the range 5-20 fm, the mass radius of 2.42 fm and the charge radius of 2.36 fm – the determination of the expressions for these radiuses are given, for example, in papers (Dubovichenko 2012a; Dubovichenko & Dzhazairov-Kakhramanov 1994, 1997). Cited above AC errors are determined by their averaging over the specified range of distances. The elastic scattering phase shift of such potential smoothly decreases and at 1.0 MeV has a value of about 174(1)°.

This potential of the GS does not have the FS in full accordance with the classification of states according to Young tableaux, given above. The GS potential parameters were chosen exclusively for the correct description of the obtained AC value that is equal to 1.39. However, as we have seen the ANC and the $S$ have sizable errors, therefore let us consider another two variants of the GS potentials that lead to the similar results.

The first of them has a smaller width and parameters

$$V_{3/2} = 71.714957 \text{ MeV and } \gamma_{3/2} = 0.2 \text{ fm}^{-2}. \qquad (5)$$

and leads to the same binding energy of -4.063900 MeV with an accuracy of $10^{-6}$ MeV (Dubovichenko 2012b), the AC equals of 1.31(1) on the interval 5-20 fm, the mass radius of 2.41 fm and the charge radius of 2.35 fm. The elastic scattering phase shift of this potential gradually decreases and at 1.0 MeV has the value of 174(1)°.

One more potential of the GS is wider than previous one and has the parameters

$$V_{3/2} = 56.827345 \text{ MeV and } \gamma_{3/2} = 0.15 \text{ fm}^{-2}. \qquad (6)$$

It allows to get the same binding energy of -4.063900 MeV with an accuracy of $10^{-6}$ MeV (Dubovichenko 2012b), the AC equals of 1.56 (1) on the interval of 5-18 fm, the mass radius of 2.44 fm and the charge radius of 2.36 fm. The elastic scattering phase shift gradually decreases and at 1.0 MeV has the value of 173(1)°. All these three potentials, in general, reflect the first interval errors of determination the dimensionless $C_W = 1.39(15)$, and the potential (6) agrees also with the latest estimation of the AC value equals $C_W = 1.54(58)$.

The potential of the first excited $^4P_{1/2}$ state is obtained by the simple decreasing of the depth of the GS potential (4) and has the parameters

$$V_{1/2} = 56.727582 \text{ MeV and } \gamma_{1/2} = 0.18 \text{ fm}^{-2}. \qquad (7)$$

The potential leads to the binding energy of -1.372900 MeV with an accuracy of $10^{-6}$ MeV (Dubovichenko 2012b), the AC is equal to 0.80(1) on the interval 5-28 fm, the mass radius of 2.53 fm and the charge radius of 2.37 fm. The value of the AC is not significantly greater than the value $C_W = 0.77$, obtained in paper (Huang, Bertulani & Guimaraes 2010). The elastic scattering phase shift gradually decreases and at 1.0 MeV has a value of 168 (1)°.

There is another variant of the FES potential, obtained on the basis of (5)

$$V_{1/2} = 62.433328 \text{ MeV and } \gamma_{1/2} = 0.2 \text{ fm}^{-2}. \qquad (8)$$

The potential gives the binding energy of -1.372900 MeV with an accuracy of $10^{-6}$ MeV (Dubovichenko 2012b), the AC is equal to 0.76 (1) on the interval 5-30 fm, the mass radius of 2.51 fm and the charge radius of 2.37 fm. For this potential the value of the AC practically coincides with the results of paper (Huang, Bertulani & Guimaraes 2010). The elastic scattering phase shift gradually decreases and at 1.0 MeV has a value of 169(1)°.

Now let us consider the potential construction criteria for the $^4S$ scattering wave. First of all, as it was indicated above, such potential must have the forbidden state, which, however, should not necessarily be bound. Furthermore, since there are no results of the phase analysis of the n$^8$Li elastic scattering, and in the spectra of $^9$Li at the energy below 1.0 MeV there are no resonances of the positive parity, we assume that the $^4S_{3/2}$ potential should lead in this energy range to almost zero phase shifts.

The potential of the nonzero depth with the FS should be relatively narrow, to obtain a smoothly varying scattering phase close to zero. The parameters of this potential can have the form

$$V_S = 327.0 \text{ MeV and } \gamma_S = 1.0 \text{ fm}^{-2}. \tag{9}$$

The potential leads to the scattering phase shifts that is shown in Fig. 1 by the solid line, which at the energies up to 1.0 MeV is located in the range from +0.03° to -0.12°.

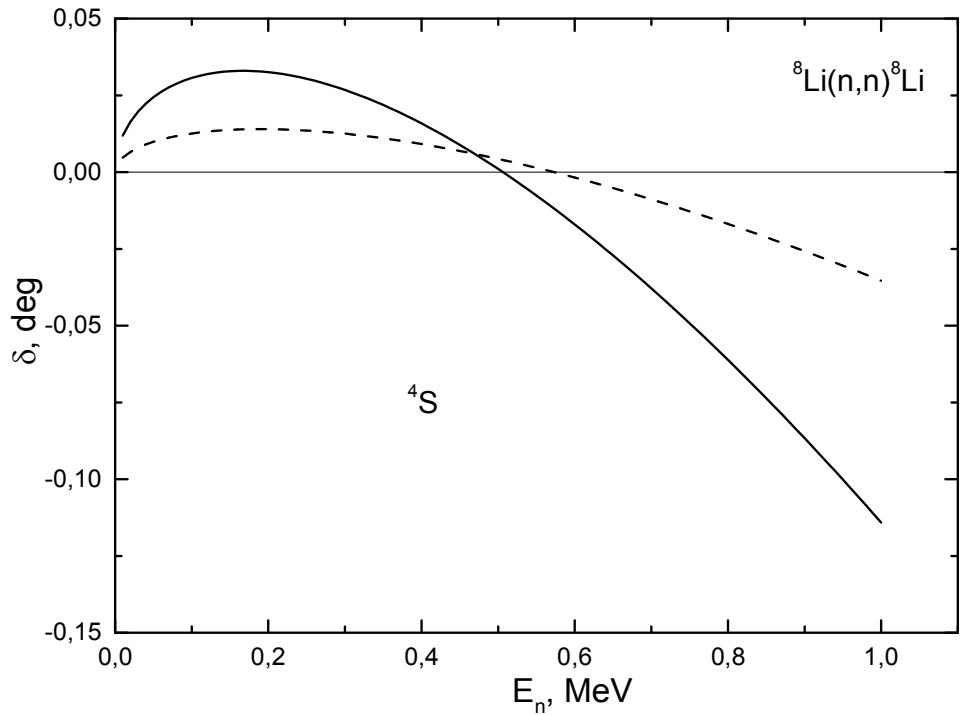

Fig. 1. The phase shift of the elastic n$^8$Li scattering in the $^4S$ wave at low energies. Lines are explained in the text.

The second variant of the scattering potential has the parameters,

$$V_S = 653.5 \text{ MeV and } \gamma_S = 2.0 \text{ fm}^{-2}. \tag{10}$$

and its phase is represented in Fig. 1 by the dashed line and at the considered energies has the value less than ±0.05°.

In addition to these two variants of the potentials of the $^4S$ scattering, furthermore the variant of such interaction with zero depth will be considered, which leads to a zero phase shifts and has no FS.

The potential without the bound forbidden state for the $^4P_{5/2}$ resonance scattering has the following parameters

$$V_P = 54.446 \text{ MeV and } \gamma_P = 0.2 \text{ fm}^{-2}. \tag{11}$$

and leads to the resonance energy of 0.261(1) MeV (l.s.) at the width $\Gamma_{cm} = 106(1)$ keV with a scattering phase shift of 90.0°(1), which is in agreement with the data of (Tilley et al. 2004).

# 7. Total cross sections of the neutron radiative capture on $^8$Li

As already mentioned, we will assume that the radiative capture for the $E$1 process comes from the $^4S_{3/2}$ scattering wave to the $^4P_{3/2}$ GS of $^9$Li in the n$^8$Li channel (Dubovichenko & Dzhazairov-Kakhramanov 2012a; Dubovichenko 2014). The calculations of the total cross sections for the potential of the GS (4) made by us leads to the results shown in Fig. 2 by the solid line. The results for the potential of the GS (5) are shown by the dashed line and for the potential (6) by the dotted line. In all of these calculations for the potential of the elastic scattering the parameters (9) were used.

As we can see from these results, the shape of the cross sections weakly depends on the potential of the GS and all the proposed variants reproduce the available experimental data of paper (Zecher 1998) acceptably. For comparison, the results for the potential of the $^4S$ scattering wave of zero depth are given, i.e., without the FS, and for the potential of the GS (4) – blue dot-dashed line in Fig. 2. They do not significantly differ from the results of calculation of the cross sections for the scattering potential (9) with the same interaction of the GS (4), which are shown by the solid line.

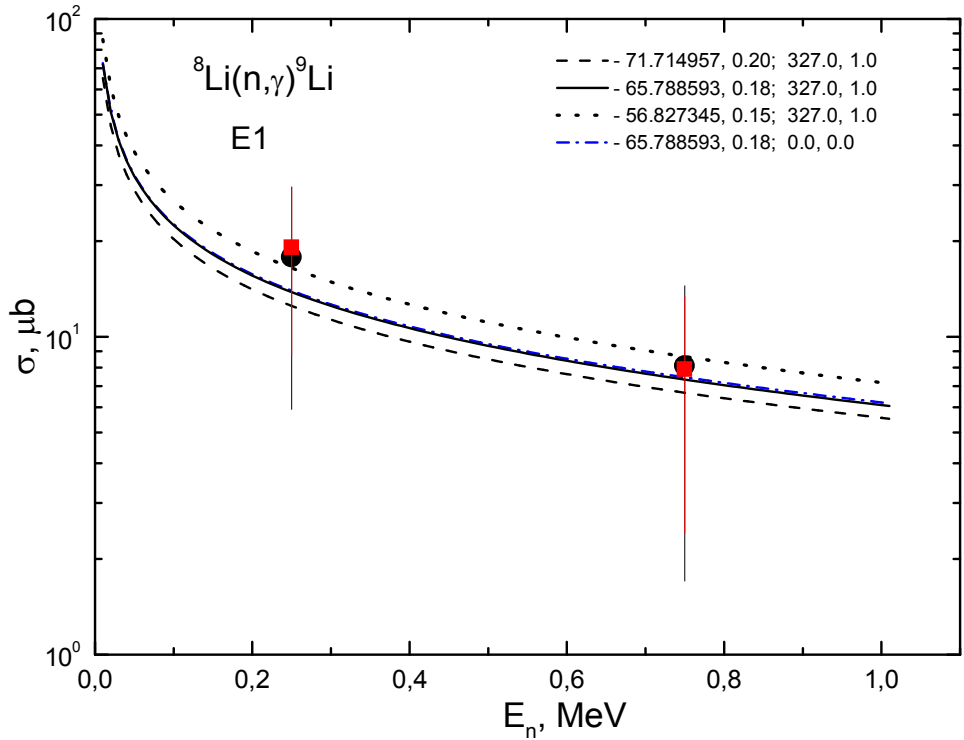

Fig. 2. The total cross sections of the $^8$Li(n,γ)$^9$Li radiative capture to the GS of $^9$Li. Black points – exp. data from (Zecher 1998) on Pb, red squares – data from (Zecher 1998) on U. Lines are explained in the text.

Now we present the results of calculations of the total cross sections of the radiative capture for the scattering potential (10). All of them are shown in Fig. 3 – all lines are marked similarly in Fig. 2. It can be seen from these results that the total capture cross section not only weakly depends on the potential of the ground state, but practically do not depend on the shape and depth of the scattering potential, and such potential does not necessarily has the bound FS – the equality of the phase shifts to zero is only important.

The results of calculation of the cross sections of work (Banerjee Chatterjee & Shyam 2008) are shown in Fig. 3 by the green dot-dashed line, which are based on the Coulomb dissociation of $^9$Li. Let us note, that results in (Zecher 1998) are averaging over the energy ranges 0 – 0.5 and 0.5 – 1.0 MeV. In addition, in paper (Zecher 1998) only the upper limit of the capture cross sections is given, therefore the results of (Banerjee Chatterjee & Shyam 2008) describe such experimental data quite accurately. Furthermore, the results of (Descouvemont 1993), based on a microscopic cluster model, are shown by the blue dot-dot-dashed line in Fig. 3. These results do not differ significantly from our own, shown in Figs. 2 and 3 for the potential of the GS (5) by the dashed line.

Our calculations of the $E1$ cross section for the GS potential (5) and the scattering potential (10) at thermal energy of 25.3 meV leads to the value of 41.3 mb. The results of this calculation, as the function of energy in the region of 25 meV – 1.0 MeV are shown in Fig. 4 by the solid line. For the same GS potential and the scattering potential (9) the thermal cross section has the value of 41.2 mb and for the scattering potential of the zero depth without FS the value of 41.4 mb was obtained. Let us note that, for example, in paper (Descouvemont 1993) for the total cross section value at thermal energy the value of 37.9 mb is given.

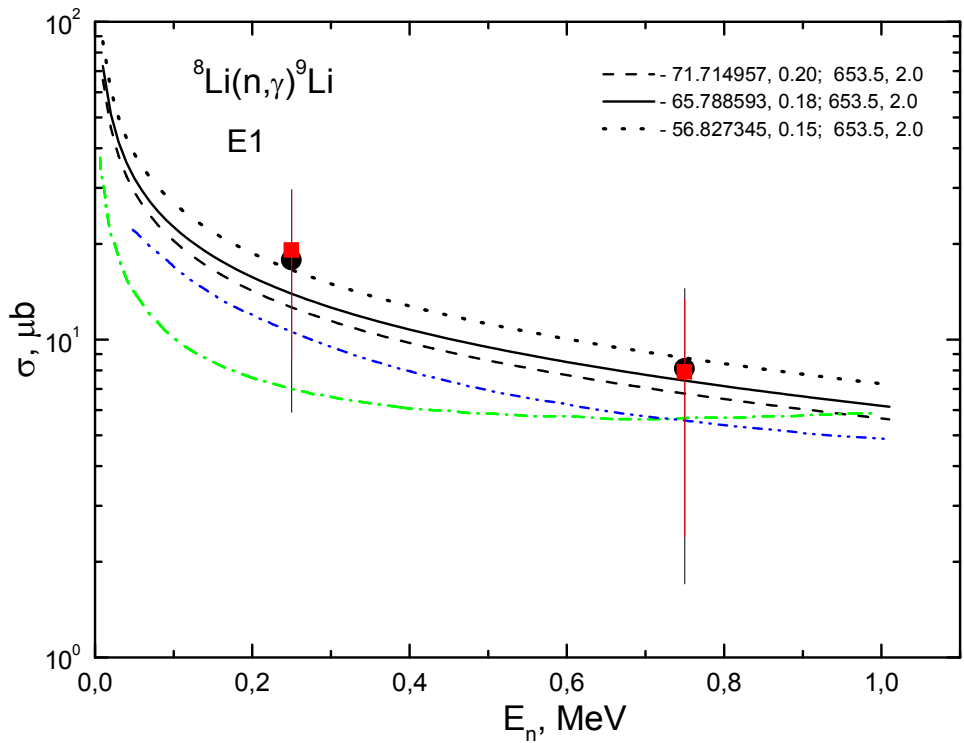

Fig. 3. The total cross sections of the $^8$Li(n,γ)$^9$Li radiative capture to the GS of $^9$Li. Black points – exp. data from (Zecher 1998) on Pb, red squares – data from (Zecher 1998) on U. Lines are explained in the text.

The $E1$ transition cross section to the FES with the potential (8) and the scattering potential (10) is shown in Fig. 4 by the dashed line and has a much smaller value of 4.5 mb at thermal energy of 25.3 meV. Rather greater value of 4.9 mb was obtained for the FES potential (7) and the scattering potential (10). This cross section has the values of almost an order of magnitude smaller than that for the transition to the GS at the range of 0.1–1.0 MeV.

The cross section of the $M$1 transition $^4P_{5/2} \to {}^4P_{3/2}$ from the resonance scattering wave with the potential (11) to the GS with the potential (5) is shown by the dotted line in Fig. 4. Blue dash-dot line with the resonance in the area of 0.26 MeV is the summed total cross section of the $E$1 and $M$1 transitions to the GS. It can be seen from these results that the consideration of the $M$1 transition leads to a small resonance in the cross sections, which at so large measurement errors does not affect on their magnitude significantly. Additional consideration of the $E$1 transition to the $^4P_{1/2}$ FES increases the summed total cross sections approximately on 10%, leading to thermal cross sections of about 46 mb.

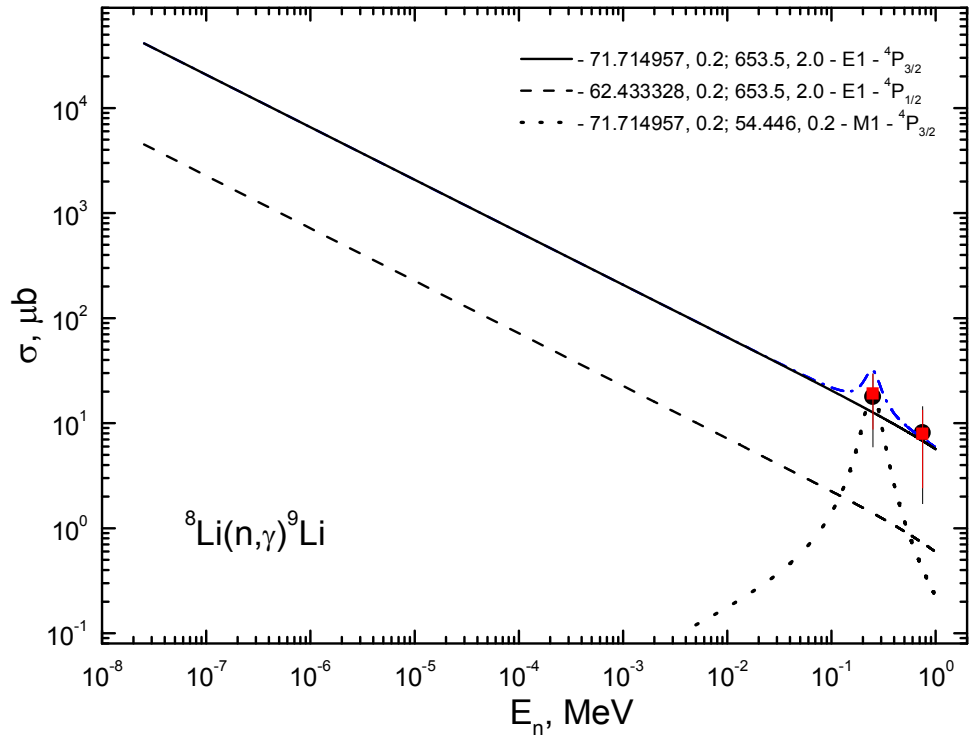

Fig. 4. The total cross sections of the $^8$Li(n,γ)$^9$Li radiative capture to the GS of $^9$Li. Black points – exp. data from (Zecher 1998) on Pb, red squares – data from (Zecher 1998) on U. Lines are explained in the text.

Since, previously we have obtained all potentials of the $^4P$ waves, which can be used for calculation of the characteristics of the BSs and for calculation the phase shifts of the elastic scattering, they can be used for consideration of the $E$2 transitions from the scattering states to the GS:

No. 1. $^4P_{1/2} \to {}^4P_{3/2}$
No. 2. $^4P_{3/2} \to {}^4P_{3/2}$
No. 3. $^4P_{5/2} \to {}^4P_{3/2}$

And also for the $E$2 capture to FES

No. 4. $^4P_{3/2} \to {}^4P_{1/2}$

No. 5. $^4P_{5/2} \to {}^4P_{1/2}$

The calculation results of the total cross sections for the $E2$ transitions from all $^4P$ scattering waves to the GS are shown in Fig. 5 by the red dot-dashed line. Results for the process No. 1 with the scattering potential in the $^4P_{1/2}$ wave (8) and the GS potential (5) are presented by the dotted line. For the transition No. 2 with the GS potentials (5) in the continuous and discrete spectrum – by the dashed line. For the process No. 3 with the $^4P_{5/2}$ scattering potential (11) and the GS potential (5) – by the solid line.

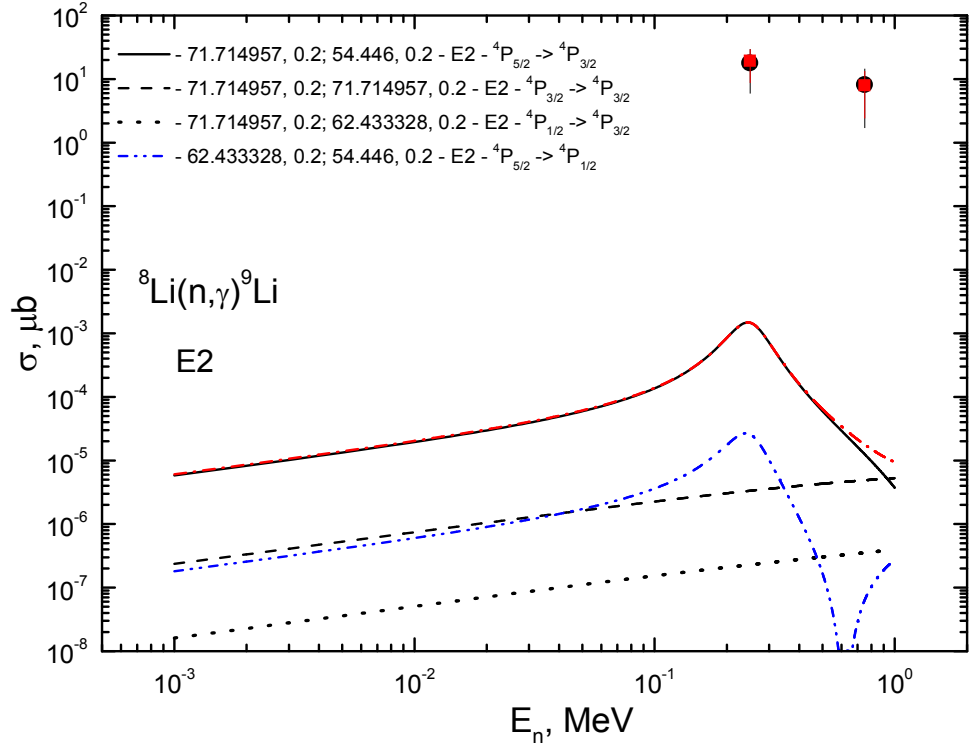

Fig. 5. The total cross sections of the $^8$Li(n,γ)$^9$Li radiative capture for the $E2$ transitions from all $^4P$ scattering waves to the GS of $^9$Li. Black points – experimental data from (Zecher 1998) on Pb, red squares – data from (Zecher 1998) on U. Lines are explained in the text.

The total cross sections for transitions to the FES have an even smaller value for the resonance process No. 5 with the scattering potential (11) and the FES potential (8) are shown in Fig. 5 by the blue dot-dot-dashed line. It can be seen from these results that in the used MPCM with the potentials proposed above the considered $E2$ transitions, even from the resonance $^4P_{5/2}$ wave, actually do not give any contribution to the total cross sections of the neutron radiative capture on $^8$Li.

Since at the energies from 25.3 meV and up to, approximately, 100 keV the calculated cross section is a straight line (the solid line in Fig. 4), it can be approximated by a simple function of energy of the form (Dubovichenko 2014):

$$\sigma_{ap} (\mu b) = \frac{A}{\sqrt{E_n (\text{keV})}}. \tag{12}$$

The constant value of $A = 207.8550$ μb keV$^{1/2}$ was determined by one point in the calculated cross sections at the minimal energy of 25.3 meV. The modulus of relative deviation

$$M(E) = \left|[\sigma_{ap}(E) - \sigma_{theor}(E)] / \sigma_{theor}(E)\right|$$

between the theoretically calculated cross section ($\sigma_{theor}$) and its approximation ($\sigma_{ap}$) given above by the function (12) in the energy range up to 10 keV has a value of less than 0.2%, while at the increasing of the energy up to 100 keV is increasing to 1.5–2.0%.

It is realistic to assume that this form of dependence of the total cross section from the energy will be survived at lower energies. Therefore, it is possible to do the estimation of the cross section, for example, at the energy of 1 μeV (1 μeV = 10$^{-6}$ eV), which gives the value of the order of 6.6 b on the basis of the given above expression for approximation of the cross section (12).

If one uses the zero potential of scattering for calculation of the cross section, then the coefficient in expression (12) for the approximation of the calculation results of the total cross section is equal to $A = 208.1156$ μb keV$^{1/2}$. The estimation of accuracy of approximation of the calculated capture cross section by the function (12) in this case is on the same level.

## 8. Conclusion

Thus, in the frame of the MPCM it is possible to construct the two-body potentials of the n$^8$Li interaction, which allow one to describe the available experimental data for the total cross section of the neutron radiative capture on $^8$Li at low and ultralow energies correctly. Theoretical cross sections are calculated from the thermal energy 25.3 meV to 1.0 MeV and approximated by a simple function of energy, which can be used for calculation of the cross sections at energies up to 50–100 keV. The proposed variants of the GS potentials of $^9$Li in the n$^8$Li channel allow one to obtain the AC in the limit of the existent for it errors and lead to a reasonable description of the $^9$Li radius. Any of the proposed variants of the GS potentials allow one to get the calculated capture cross sections that reasonable agree with the available experimental data (Zecher 1998). The calculation results of the cross sections are not sensitive to the number of the bound FSs and the width of the $^4S$ scattering potential. The obtained results for the total cross sections and static characteristics of $^9$Li depend only on the parameters of the GS potential of this nucleus in the n$^8$Li channel.

This shows that due to the large clusterization of $^9$Li in the n$^8$Li channel the simple two-body potential cluster model independently of the presence of FSs in the $^4S_{3/2}$ wave allows to describe the main characteristics of $^9$Li correctly, as a weakly connected n$^8$Li system in the discrete spectrum (the binding energy, radiuses, the asymptotic constant for the n$^8$Li channel) and the total cross section of the neutron capture on $^8$Li. Many of the earlier results (Dubovichenko 2012a, 2014), for most of the above considered reactions, depended significantly on the presence of the bound FSs in the different partial waves. In this case, the independence of the results for the total cross sections from the presence of the bound FS in the $^4S_{3/2}$ wave, which follows from the classification of states according to Young tableaux, apparently, can be regarded as proof of an exceptionally large clusterization of particles to the n$^8$Li channel in $^9$Li.

As mentioned, we have previously considered 27 radiative capture reactions (Dubovichenko 2012a, 2013d, 2014; Dubovichenko & Dzhazairov-Kakhramanov 2012b, 2012c, 2013, 2014; Dubovichenko & Uzikov 2011; Dubovichenko, Dzhazairov-Kakhramanov & Afanasyeva 2013; Dubovichenko, Dzhazairov-Kakhramanov & Burkova 2013) the total cross sections of which can be described within the framework of the MPCM and the given process is already the 28$^{th}$ reaction of this type. And in this case it is possible to get a quite reasonable description of the known data on the static characteristics of $^9$Li and the total cross sections of the neutron radiative capture on $^8$Li.

## Acknowledgments

This work was supported in the framework of the financing program "Studying of the thermonuclear processes in the Universe" of the Ministry of Education and Science of the Republic of Kazakhstan through the V.G. Fessenkov Astrophysical Institute, NCSRT RK.

In conclusion, the authors express their deep gratitude to Strakovsky I.I. (GWU, Washington, USA), Uzikov Y.N. (JINR, Dubna, Russia) and Ibraeva E.T. (INP, Almaty, Kazakhstan) for a discussion of certain issues touched in the paper.